\documentclass[]{emulateapj}
\usepackage{natbib}

\shorttitle{STIS spectroscopy of RZ~2109}
\shortauthors{Peacock et al.}

\begin{document}

\title{Spatially resolved spectroscopy of the globular cluster RZ~2109 and the nature of its black hole$^{\dag}$}

\author{Mark B. Peacock$^{1}$, Stephen E. Zepf$^{1}$, Arunav Kundu$^{2}$, Thomas J. Maccarone$^{3}$, Katherine L. Rhode$^{4}$, John J. Salzer$^{4}$, Christopher Z. Waters$^{5}$, Robin Ciardullo$^{6, 7}$, Caryl Gronwall$^{6, 7}$, Daniel Stern$^{8}$}
\affil{$^{1}$Department of Physics and Astronomy, Michigan State University, East Lansing, MI 48824, USA; mpeacock@msu.edu}
\affil{$^{2}$Eureka Scientific, Inc., 2452 Delmer Street, Suite 100 Oakland, CA 94602, USA}
\affil{$^{3}$School of Physics and Astronomy, University of Southampton, Southampton, SO17 1BJ, UK}
\affil{$^{4}$Department of Astronomy, Indiana University, Bloomington, IN 47405, USA}
\affil{$^{5}$Institute for Astronomy, University of Hawaii at Manoa, Honolulu, HI 96822, USA}
\affil{$^{6}$Department of Astronomy and Astrophysics, The Pennsylvania State University, University Park, PA, 16802, USA}
\affil{$^{7}$Institute for Gravitation and the Cosmos, The Pennsylvania State University, University Park, PA 16802, USA}
\affil{$^{8}$Jet Propulsion Laboratory, California Institute of Technology, Pasadena, CA 91109, USA}

\email{$^{\dag}$Based on observations made with the NASA/ESA Hubble Space Telescope, obtained from the data archive at the Space Telescope Science Institute. STScI is operated by the Association of Universities for Research in Astronomy, Inc. under NASA contract NAS 5-26555}

\begin{abstract}
\label{sec:abstract}

We present optical {\it HST}/STIS spectroscopy of RZ~2109, a globular cluster in the elliptical galaxy NGC~4472. This globular cluster is notable for hosting an ultraluminous X-ray source as well as associated strong and broad [O~{\sc iii}]~$\lambda\lambda$4959,~5007 emission. We show that the {\it HST}/STIS spectroscopy spatially resolves the [O~{\sc iii}] emission in RZ~2109. While we are unable to make a precise determination of the morphology of the emission line nebula, the best fitting models all require that the [O~{\sc iii}]~$\lambda$5007 emission has a half light radius in the range 3-7 pc. The extended nature of the [O~{\sc iii}]~$\lambda$5007 emission is inconsistent with published models that invoke an intermediate mass black hole origin. It is also inconsistent with the ionization of ejecta from a nova in the cluster. The spatial scale of the nebula could be produced via the photoionization of a strong wind driven from a stellar mass black hole accreting at roughly its Eddington rate.

\end{abstract}


\section{Introduction}
\label{sec:intro}

Some of the best evidence for black holes (BHs) in globular clusters (GCs) has come from X-ray studies of extragalactic GCs. Specifically, X-ray sources with luminosities in excess of the Eddington limit for an accreting neutron star have been observed -- suggesting accretion on to a more massive BH primary \citep[e.g.,][]{Sarazin00,Angelini01,Fabbiano06}. Some of these ultraluminous X-ray sources are observed to vary on short time scales, which is important because it eliminates the possibility that the high X-ray luminosity arises from the superposition of multiple accreting neutron stars \citep{Kalogera04}. To date, six GC X-ray sources have been proposed as accreting BHs based on these two key properties (luminosity and variability). These sources reside in GCs in the galaxies NGC~4472 \citep[two systems;][]{Maccarone07, Maccarone11a}, NGC~1399 \citep{Shih10}, NGC~3379 \citep[two systems;][]{Brassington10,Brassington12} and NGC~4278 \citep{Brassington12}. In addition to these objects, BHs have been proposed to exist in GCs in two other galaxies. A BH candidate was identified in a GC in the elliptical galaxy NGC~1399 based on non-variable ultraluminous X-ray emission and nebular emission lines \citep{Irwin10}. Nine GCs in M31 have fainter X-ray emission, but spectra that look like BH low mass X-ray binaries (LMXBs), rather than neutron star LMXBs \citep{Barnard08, Barnard12}.

The ultraluminous and variable X-ray source XMMU122939.7+075333, located in the GC RZ~2109 in the elliptical galaxy NGC~4472, is a very strong BH candidate \citep{Maccarone07}. In a 2004 {\it XMM~Newton} observation, the peak luminosity of this source was L$_{{\rm x}}$~=~4$\times$10$^{39}$erg/s, which is an order of magnitude greater than the Eddington limit for an accreting neutron star. Furthermore, during this observation the luminosity varied by a factor 7 in only a few hours \citep{Maccarone07}. The optical spectrum of this cluster revealed extraordinary [O~{\sc iii}] emission \citep{Zepf07}. Follow-up observations confirmed the presence of extremely broad [O~{\sc iii}]~$\lambda\lambda$4959,~5007 emission lines with velocity widths of $\sim$1500~kms$^{-1}$ and a luminosity of L([O~{\sc iii}]~$\lambda$5007)~=~1.4$\times$10$^{37}$~ergs/s \citep{Zepf08}. In addition to the extreme luminosity and velocity width of this emission, the non-detection of H$\beta$ from the cluster requires that the nebula is extremely hydrogen deficient, with L([O~{\sc iii}])/L(H$\beta$)~$\gtrsim$~30. Strong forbidden lines are extremely rare in GCs \citep[e.g.][]{Peacock12} and the emission is very likely related to the (equally rare) ultraluminous X-ray source in the cluster. As noted by \citet{Zepf07}, this nebula can therefore provide important clues as to the nature of the accreting compact object in RZ~2109. In this paper, we present spatially resolved {\it Hubble Space Telescope (HST)}/ Space Telescope Imaging Spectrograph (STIS) longslit spectroscopy of RZ~2109 (Section \ref{sec:stis_data}). In Section \ref{sec:profiles}, we use these data to investigate the size of the emitting nebula and, in Section \ref{sec:models}, we compare this with the expectations from the current models that have been proposed to explain the X-ray and optical emission from RZ~2109. 

\section{STIS spectroscopy}
\label{sec:stis_data}

In order to study the spatial extent of the [O~{\sc iii}] emission from RZ~2109, we obtained {\it HST} STIS longslit spectroscopy (proposal ID 11703, PI Zepf). The cluster was observed on four dates: 2009/12/29; 2010/01/01; 2010/03/03; 2010/04/11. All observations were obtained through the 52x0.2$\arcsec$ slit using the G430L grating (with a central wavelength of 4300${\rm \AA}$). On each date 4 observations were taken, resulting in 16 exposures and a total exposure time of 41.2~ksec. 

We used the calibrated 2D rectified spectral images produced by the standard STIS pipeline routines run by {\sc calstis} (these are the images given `sx2' suffix by the pipeline). This performs the bias subtraction, cosmic ray rejection, dark subtraction and applies the flat fields. It then applies a geometric, wavelength and flux calibration. The resulting image has a scale of 1 pixel = 0.05078$\arcsec$ in the spatial direction and is wavelength calibrated in the dispersion direction, with 1 pixel = 2.73${\rm \AA}$ and an estimated line spread function of 1.5~pixels (4.1${\rm \AA})$. 

The final reduced spectral image was found to have significant numbers of hot pixels remaining. We (in combination with the STIS helpdesk) experimented with rerunning the {\sc calstis} pipeline using darks taken with similar temperatures and within the same annealing period, but found no improvements over using the master darks for our observations. While the darks vary significantly with time, the greater S/N of the master dark gave better results. We were therefore unable to correct for these pixels. Instead, we flagged the bad pixels by running a 5$\sigma$ rejection script over 10~pixel regions in the dispersion direction. 

The 16 resulting 2D spectral images were combined using the {\sc iraf} task {\sc imcombine}. The spatial and wavelength calibrations were found to be accurate enough to combine the reduced images without applying any shifts. The images were scaled and weighted based on their exposure times and combined together using mask images to reject bad pixels. The following analysis is based on this final combined image. 

The images are flux calibrated by the pipeline to give the surface brightness per ${\rm \AA}$ in units of ${\rm erg/cm^{2}/s/\AA/arcsec^{2}}$. To produce the 1D spectrum of RZ~2109, we integrated the spectrum over a region 7 pixels wide (as is recommended for STIS observations). The resulting fluxes were multiplied by the DIFF2PT keyword (produced by the pipeline) to give the flux calibrated spectrum, corrected for slit losses (assuming a point source). Given that our source is slightly extended, we experimented with wider extraction regions, but found the default resulted in the best S/N spectrum. Finally, the fluxes were corrected for known charge transfer efficiency (CTE) effects \citep[see e.g.][]{Goudfrooij06}. This signal loss is significant for these observations of a faint target, because they were taken before the extent of the CTE problem was commonly known and hence centered on the middle of the detector (rather than the now commonly used `E1' aperture, located close to the detector readout side). The CTE correction was calculated as 0.52, based on the equation defined in Section 3.4.6 of the ``STIS data handbook" \citep[v6.0,][]{Bostroem11}. Applying this correction produced the final flux calibrated spectrum. 

\begin{figure}
 \centering
 \includegraphics[height=88mm,angle=270]{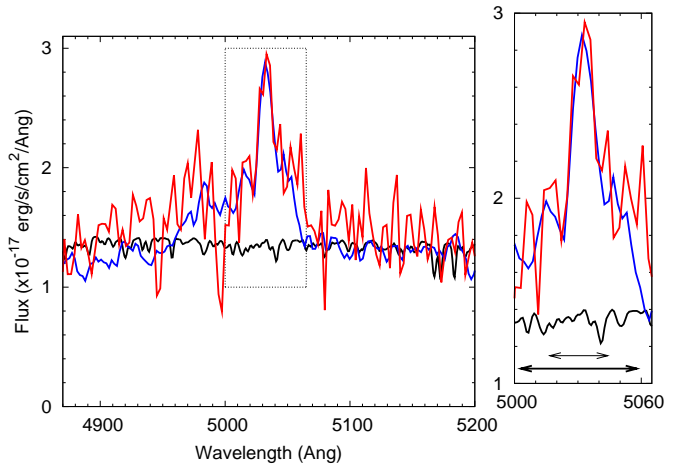} 
 \caption{The STIS spectrum of RZ~2109 (red). The blue line shows a previously obtained Keck spectrum of this cluster. The black line is a model of a 12~Gyr star cluster with the magnitude and metallicity of RZ~2109. The zoomed panel shows the 5007${\rm \AA}$ emission line region (located at 5030.2${\rm \AA}$ at the redshift of RZ~2109). The model is in good agreement with the STIS continuum level, but there is a clear excess of [O~{\sc iii}] emission. The arrows indicate a 10~pixel and 20~pixel wide region, centered on rest frame 5007${\rm \AA}$.}
 \label{fig:rz2109_spec} 
\end{figure}

Figure \ref{fig:rz2109_spec} shows the [O~{\sc iii}]~$\lambda$5007 region of the extracted STIS spectrum of RZ~2109. In this figure, we have also plotted a previously obtained Keck/LRIS spectrum of this cluster \citep{Zepf08} and a simple stellar population model from \citet{Maraston05} of a 12~Gyr cluster, with V=21.0~mag and [Fe/H]=-1.3 \citep[similar to the parameters measured for RZ~2109;][]{Steele11}. The flux of the continuum level is in very good agreement with the expectation from the cluster model, suggesting that the majority of the cluster luminosity is included in the STIS slit. The continuum of the Keck spectrum is slightly lower at shorter wavelengths. However, this spectrum is not fully flux calibrated and was only scaled to have the correct flux at 5007${\rm \AA}$, based on its V-band magnitude \citep{Zepf08}. It is therefore in good agreement with the STIS observations. Compared with the spectrum expected for a stellar population with RZ~2109's properties (black line), it can be seen in both the old (Keck, blue line) and new (STIS, red line) spectra, that there is strong and broad excess emission around 5007${\rm \AA}$ (5030.2${\rm \AA}$ at the velocity of RZ~2109). A discussion of the line strengths and any potential temporal variability in these spectra will be presented in Steele et al. (in preparation).

\section{Stellar profile of RZ~2109} 

The size, density, and mass of a GC can significantly influence its population of accreting compact objects. For example, it is now well established that the formation of LMXBs is orders of magnitude more efficient in GCs than the field of the galaxy \citep[e.g.][]{Clark75, Angelini01, Sarazin03, Kim06, Kundu07}. This is likely due to dynamical interactions enhancing the formation of LMXBs. Therefore, clusters with higher stellar interaction rates are expected to host more LMXBs, as observed \citep[][]{Verbunt87,Jordan07,Peacock09}. The presence of an intermediate mass BH (IMBH), or even stellar mass BHs, in a GC can also significantly affect the cluster's structure, potentially producing relatively large cores with shallow central cusps \citep[e.g.][]{Hurley07,Trenti07}. Given RZ~2109's prominence as one of the few clusters known to host a BH, it is interesting to consider the structure of the cluster. 

Due to the small angular size of GCs, it is observationally very challenging to measure the structure of clusters at the distance of NGC~4472 (which is taken to be 16.0~Mpc throughout this paper). However, it is possible to estimate such parameters by fitting PSF convolved King models to high spatial resolution {\it HST} images, provided the S/N of the observation is high enough \citep{Carlson01}. Such methods have previously been used to estimate the structure of GCs around other Virgo cluster galaxies \citep[e.g.][]{Jordan04, Waters07, Williams07, Madrid09}. 

\begin{figure}
 \centering
 \includegraphics[height=90mm,angle=270]{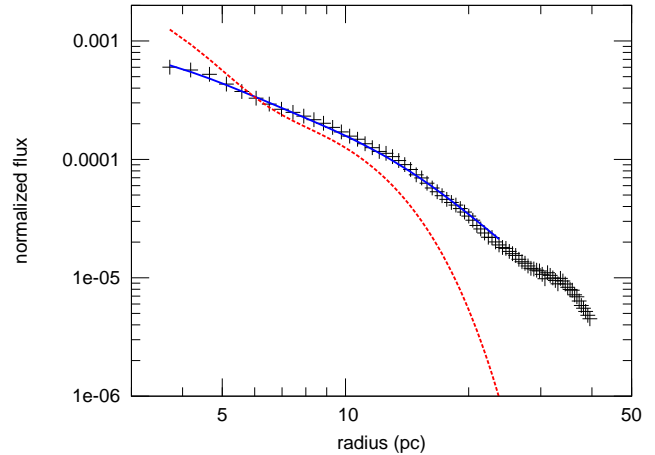} 
 \caption{Radial profile of RZ~2109, as measured from the STIS target acquisition images (black pluses). The dotted red line indicates the point spread function (PSF) of these observations and the solid blue line shows the previously determined King model fit to RZ~2109, convolved with the PSF. This King model fit provides an excellent match the new STIS images. }
 \label{fig:rz2109_prof} 
\end{figure}

The target acquisition images that were taken prior to each STIS spectrum provide some of the highest S/N and spatial resolution images available for this cluster. In total eight 240~s exposures were obtained on the STIS CCD through the F28X50LP filter (covering $5500{\rm \AA}<\lambda<10000{\rm \AA}$). The STIS pixels are slightly undersampled. We therefore drizzled the images together onto a grid with pixels 1/10 the native pixel size. Because the orientation and location of the source on the detector varied between each exposure, this process reduces pixellation effects in the final images. Before combination, all images were centered by using the {\sc iraf/stsdas} task {\sc ellipse} to determine the center of the cluster in each exposure. The resulting image has a total exposure time of 1920~s. The cluster profile, obtained from this combined image using {\sc ellipse}, is plotted in Figure \ref{fig:rz2109_prof} (black plusses). 

We compared the cluster profile to {\it HST}/WFPC2 observations of the cluster through the F555W filter (proposal ID 11209, PI Zepf). Waters et al. (in preparation) have previously fit PSF convolved King models to this WFPC2 image of the cluster using the \citet{Waters07} code {\sc superking}. The resulting structural parameters of RZ~2109 are listed in Table \ref{tab:rz2109_struct}. This King model, convolved with the PSF for these STIS observations (produced by the TinyTim PSF simulator), is plotted in Figure \ref{fig:rz2109_prof}. It can be seen that the model from the analysis of the WFPC2 images provides an excellent representation of the STIS cluster profile. Although a full discussion of the structure of this cluster is beyond the scope of this paper, the agreement of the WFPC2 model fit to the STIS data is reassuring. This is because the two datasets have different systematic uncertainties due effects such as PSF shape and pixellation, which typically dominate the uncertainties in the fits. Thus the agreement between the WFPC2 model and the STIS data provides some assurance that the parameters are reasonably well determined. In this context, we note that RZ~2109 appears to have a high concentration and a large half light and tidal radii, as may be expected given its large distance from the center of the host galaxy ($\gtrsim$30~kpc). 

\begin{table}
 {\centering
  \caption{Structural parameters of RZ~2109\label{tab:rz2109_struct}}
  \begin{tabular}{@{}cccccc@{}}
   \hline
   \hline
   \vspace{-3mm}
  \\
    M${\rm _{V}}^{\rm i}$  & W$_{0}^{\rm ii}$   & c$^{\rm iii}$     &   r$_{c}^{\rm iv}$ (pc)   &   r$_{h}^{\rm iv}$ (pc)   &   r$_{t}^{\rm iv}$ (pc)    \\
   \hline
    -9.76   & 8.4 & 1.9   &   0.92    & 6.49   &   82.2     \\
   \hline
   \end{tabular}\\
 }
\footnotesize
\vspace{1mm}
Derived from King model fits to WFPC2 observations of RZ~2109 (Waters et al., in preparation). Parameters are the $^{\rm i}$absolute V-band magnitude, $^{\rm ii}$central potential, $^{\rm iii}$concentration c~=~log($r_{t}/r_{h}$), $^{\rm iv}$core, half light and tidal radii of the cluster. \\
\end{table}

\section{Spatial distribution of the [O~{\sc iii}] nebula}
\label{sec:profiles}

\begin{figure}
 \centering
 \includegraphics[height=86mm,angle=270]{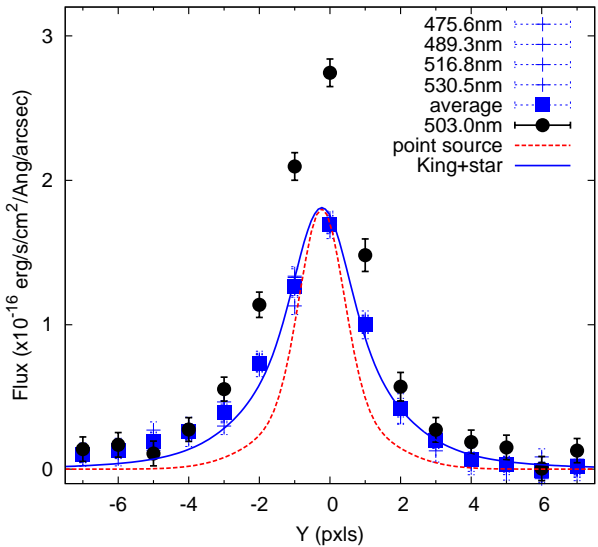} 
 \caption{The spatial profile of RZ~2109's spectrum across 50~pixel wide regions centered at 4756, 4893, 5168 and 5305${\rm \AA}$ (blue crosses). These regions are outside of the wavelength range influenced by the nebular emission and represent the cluster's continuum emission from its stellar component. The blue line represents a fit to the average profile across all four of these continuum regions. The black points show the average profile of a 20~pixel region centered on the [O~{\sc iii}] emission line. The nebular emission is centered on the cluster and very bright, increasing the cluster flux over these wavelengths by around 40$\%$. We have also plotted the estimated PSF for these observations (scaled to match the peak flux, dashed red line) and PSF convolved King model of RZ~2109 (solid blue line). It can be seen that the cluster is clearly extended and is well represented by the previously determined King model of the cluster. In figure \ref{fig:prof_oiii}, we compare the profile at 5030${\rm \AA}$ to this PSF, King model and other profiles.}
 \label{fig:prof_rz2109} 
\end{figure}

\begin{figure*}
 \centering
 \includegraphics[height=160mm,angle=270]{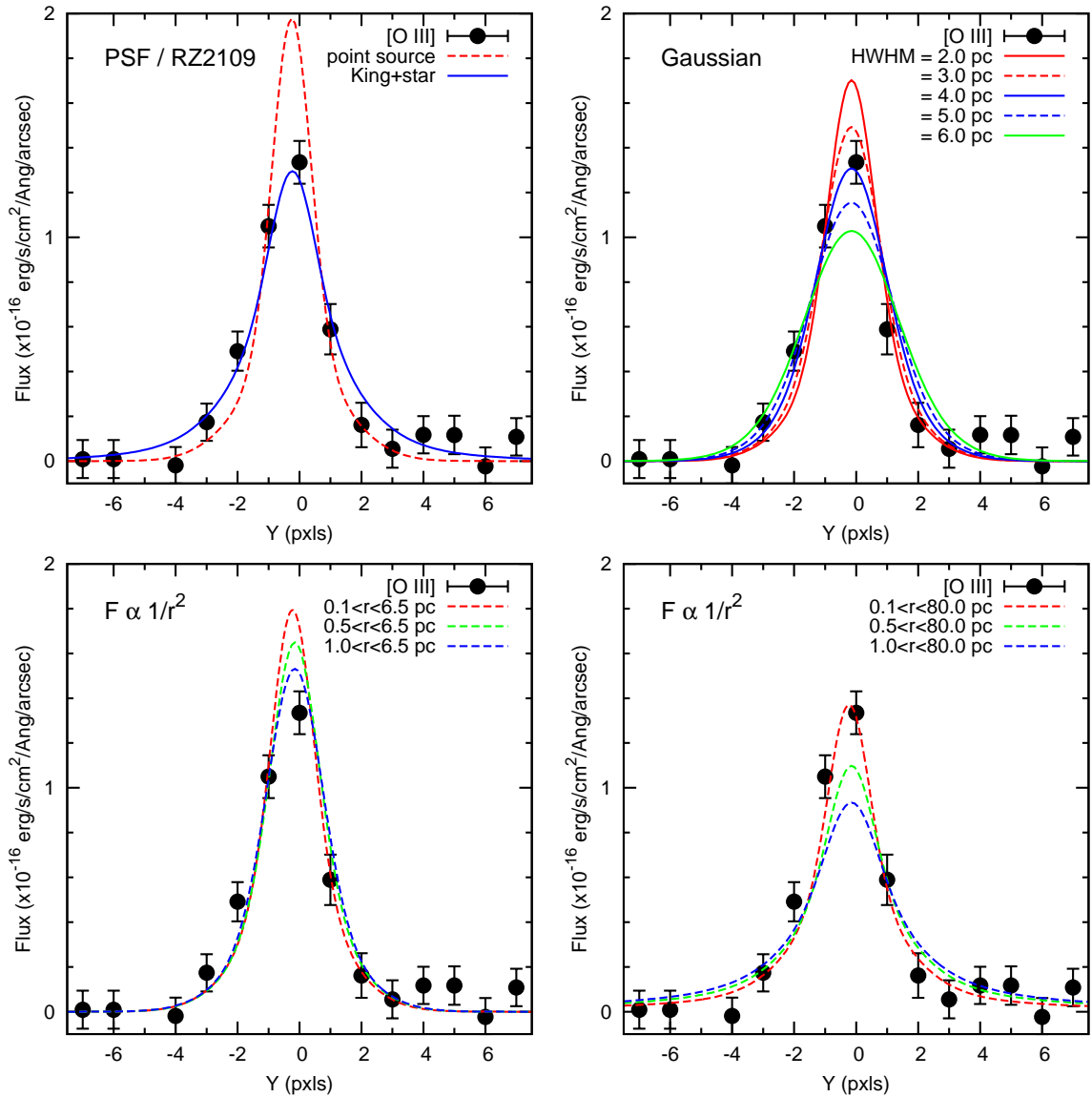} 
 \caption{Profile of the [O~{\sc iii}] emission line region (black points). The different panels compare the observed profile with those expected from: a stellar source (dashed red line, top left); the PSF convolved King model representing RZ~2109's stellar component (blue line, top left); PSF convolved Gaussians (top right); and $1/r^{2}$ profiles with varying inner and outer boundaries (bottom panels). All of the models are scaled to have the same integrated flux as the data. The reduced $\chi^{2}$ values from fitting these profiles to the data are listed in Table \ref{tab:chisq_fits}.}
 \label{fig:prof_oiii} 
\end{figure*}

At the distance of NGC~4472 (16~Mpc), RZ~2109 is spatially resolved by the {\it HST} STIS (with 1~pixel~$\sim$~3.9~pc). The resulting long slit spectrum therefore allows us to study the cluster's spectrum as a function of radius from its center. Of particular interest is the size of the [O~{\sc iii}] emitting nebula. From Figure \ref{fig:rz2109_spec}, we estimate that the emission from the [O~{\sc iii}] doublet influences the cluster spectrum over a broad wavelength range of $4940 \lesssim \lambda_{{\rm[OIII]}} \lesssim 5070{\rm \AA}$. At other wavelengths, the spectrum is consistent with that expected of a GC with the luminosity and metallicity of RZ~2109 \citep{Maraston05, Steele11}. In Figure \ref{fig:prof_rz2109}, we plot the width of the spectrum (averaged over 50~pixels = 136~${\rm \AA}$) in four regions centered at 4756, 4893, 5168 and 5305${\rm \AA}$. These regions are at wavelengths above and below the observed nebular emission and therefore represent emission from only the stellar component of the cluster. We note that, while the PSF is known to increase with increasing wavelength, the PSF and cluster profile do not vary significantly over this relatively narrow wavelength range. The dashed red line in this figure is the profile of a stellar source, the {\it HST} standard star BD+75D325. This spectrum was obtained under a similar setup to RZ~2109's and is representative of the PSF of our observations. The solid blue line is the King model profile for this cluster, produced from fits to WFPC2/STIS photometry and convolved with the stellar PSF. The cluster is clearly resolved by these observations and the profile of the spectrum is consistent with the previously determined King model representation of the cluster.

To investigate the nebula in RZ~2109, we focus on a 20~pixel (55${\rm \AA}$) region centered on the main 5007${\rm \AA}$ peak. This region includes nearly all of the 5007${\rm \AA}$ flux (see Figure \ref{fig:rz2109_spec}). We also studied 27${\rm \AA}$ and 14${\rm \AA}$ regions, which contain less of the broad [O~{\sc iii}] emission, but have a higher ratio of line to cluster flux. The profiles for the different width regions were found to be consistent. In Figure \ref{fig:prof_oiii}, we plot the profile of this nebular emission. This is the average flux over the 20~pixel [O~{\sc iii}] region minus the cluster stellar component, taken to be the average of the continuum 100~pixels either side of the line region. The cluster component at these wavelengths is expected to be relatively flat and have no emission or absorption lines with significant flux \citep[e.g.,][]{Maraston05}. Therefore, this subtracted profile should accurately represent the nebular emission. In this figure, we compare the profile with several model profiles. All of these models have been scaled to have the same integrated flux as the data. The reduced $\chi^{2}$ values from fitting a selection of these flux profiles to the data are listed in Tabel \ref{tab:chisq_fits}. 

\begin{table}
 {\centering
  \caption{Reduced $\chi^{2}$ values for a selection of flux profiles\label{tab:chisq_fits}}
  \begin{tabular}{@{}lcc@{}}
   \hline
   \hline
   \vspace{-3mm}
  \\
  Model $^{{\rm i}}$ & Parameters & $\chi^{2}/\nu$ $^{\rm ii}$ \\
   \hline
  Stellar psf  &   & 4.81\\
  Gaussian   & hwhm=2.0~pc & 2.20\\
      & hwhm=2.5~pc & 1.55\\
      & hwhm=3.0~pc & 1.13\\
      & hwhm=3.5~pc & 0.94\\
      & hwhm=4.0~pc & 0.96\\
      & hwhm=4.5~pc & 1.17\\
      & hwhm=5.0~pc & 1.52\\
      & hwhm=5.5~pc & 1.99\\
      & hwhm=6.0~pc & 2.55\\
      & hwhm=6.5~pc & 3.17\\
  F $\propto$ 1/r$^{2}$ & 0.1$<$r$<$ 6.5~pc & 2.28\\
    &   0.1$<$r$<$20.0~pc & 1.02\\
    &   0.1$<$r$<$80.0~pc & 1.23\\
    &   0.5$<$r$<$ 6.5~pc & 1.82\\
    &   0.5$<$r$<$20.0~pc & 1.33\\
    &   0.5$<$r$<$80.0~pc & 2.69\\
    &   1.0$<$r$<$ 6.5~pc & 1.26\\
    &   1.0$<$r$<$20.0~pc & 1.92\\
    &   1.0$<$r$<$80.0~pc & 4.11\\
  King model $^{{\rm iii}}$ & as in Table \ref{tab:rz2109_struct} & 1.58\\
   \hline
   \end{tabular}\\
 }
\footnotesize
\vspace{1mm}
$^{{\rm i}}$All models for the emission profile have been convolved with the PSF for these observations.
$^{{\rm ii}}$It should be noted that only the background level is fit during these comparisons (the flux is fixed to that observed and the parameters defining the shapes of the profiles are fixed at the values quoted). We therefore take the number of degrees of freedom for each model, $\nu$~=~6. 
$^{{\rm iii}}$Representing the stellar profile of RZ~2109. \\
\end{table}

It can be seen that the nebula is best represented as an extended source. The top left panel of Figure \ref{fig:prof_oiii} shows that the point source representation (dashed red line) has a narrower peak than the data and provides a poor fit (with a reduced $\chi^{2}=4.8$, rejecting the fit with a probability 99.99$\%$). The solid blue line in this panel shows a PSF convolved King model, which represents the cluster's stellar component. The nebular emission is quite well represented by this King model, suggesting that the nebula has a similar scale to the cluster's stellar component (i.e., a half light radius of 6.5~pc). However, we do not know the functional form of the [O~{\sc iii}] spatial distribution a priori, therefore we consider several other forms for the radial profile. In the top right panel of Figure \ref{fig:prof_oiii}, we compare the nebular emission to PSF convolved Gaussian functions of different widths. As can be seen from Table \ref{tab:chisq_fits}, the emission is best represented by Gaussians that have a half width half maximum (HWHM) of 4$\pm$1.5~pc. We also consider, in the bottom panels of Figure \ref{fig:prof_oiii}, the profile produced if the nebula emission followed an inverse square law from the center of the cluster. As an example, we consider three groups of models which have density profiles which truncate at 6.5, 20.0 and 80~pc. For each of these models, we start the emission at three different inner radii (0.1, 0.5, 1.0~pc). The models truncated at large radii still have significant flux at larger distances. While our data show no evidence for such emission, the errors on the measured fluxes do not allow us to discard such large scale faint emission. It can be seen that inverse square law models which have small inner radii and large outer radii provide a reasonable fit the observations. Similarly, the observations can be represented by smaller outer radii if the emission is suppressed, by some mechanism, to relatively large inner radii. Due to this degeneracy, we are unable to determine the exact morphology of the cluster -with several of the models providing a good fit. However, all of the models which provide a good fit to the data have half light radii $\sim$3-7~pc. 

The actual nebula may be more complex than the models considered \citep[e.g.,][]{Steele11}. If it were produced by winds from a central BH-LMXB \citep[as has been proposed, e.g.,][]{Zepf07}, then these winds may not produce a spherically symmetric gas distribution. Similarly LMXBs, particularly BH LMXBs \citep{King96}, can have highly variable accretion rates. The gas expelled during such accretion is likely proportional to the accretion rate, potentially producing a nebula with highly variable density as a function of radius. The data do not allow us to investigate such details, so we do not consider more complicated models. Instead, we focus on the basic constraints provided by these observations -- that the nebula is clearly extended on parsec scales and best represented by having half of its emission on scales of $\sim$5$\pm$2~pc.

\section{The [O~{\sc iii}] nebula and nature of RZ~2109's BH} 
\label{sec:models}

Since the discovery of the ultraluminous X-ray source and broad [O~{\sc iii}] emission in RZ~2109, several processes have been suggested as the possible source. It is thought that the strong and variable X-ray emission requires an accreting BH origin. However, the nature of this BH hole remains uncertain, with both stellar mass BHs and IMBHs proposed. While all current models propose that the observed nebula is ionized by the central accreting source, several models of the distribution of the nebular gas have been proposed. The spatial scale of the [O~{\sc iii}] emission, as measured from these observations, can help to distinguish between these models. 

\subsection{Ionization of a wind from a stellar mass BH}

The [O~{\sc iii}] emission observed in RZ~2109 may result from the photoionization of a wind driven from the central accreting object \citep{Zepf08}. BHs, accreting at rates near the Eddington limit, are thought to drive strong winds from the systems \citep[e.g.,][and references therein]{Proga07, Zepf08}. Such a wind would then be photoionized by the extremely bright radiation that is observed from the central source. \citet{Zepf08} demonstrated that this model is capable of producing both the luminosity and width of the observed [O~{\sc iii}] emission. However, driving this strong wind from the system requires that the compact object be accreting at a rate similar to its Eddington limit. Therefore, at the X-ray luminosities observed, this model predicts accretion on to a stellar mass BH. 

The velocity of the outflowing wind is measured to be $\sim10^{3}$~kms$^{-1}$, which is far greater than the escape velocity of a typical cluster of this size, a few~$\times10$~kms$^{-1}$. Therefore, the size of the interstellar gas is simply given by this wind velocity times the time that the X-ray binary has been accreting and driving the wind. Unfortunately, this timescale is poorly constrained. We can estimate a minimal timescale, based on X-ray detections of this source during outburst. The source was detected by {\it ROSAT} in June/July 1996 , {\it XMM-Newton} in June 2002 and January 2004, marginally by {\it Chandra} in February 2010 \citep[a complete list of X-ray detections of RZ~2109 prior to 2010 is presented in][]{Maccarone10} and was still detected in the last observation in 2011 (Tana Joseph, private communication). We therefore assume that the source has been in outburst throughtout this period, with its obsevered luminosity peaking at $\sim$4$\times10^{39}$erg/s and reducing to $\sim$1$\times10^{38}$erg/s. Assuming steady accretion over 14~yrs would suggest that the emitted gas has a minimum size of $\sim$0.05~pc. Setting an upper limit is much harder because the evolution of X-ray binaries is highly variable and uncertain. Some sources can be persistent for long timescales, while others are highly variable and switch between periods of quiescence and outburst. XMMU122939.7+075333's history prior to the {\it ROSAT} observation is therefore unknown. Given the broad range of possible accretion (and hence wind) timescales, this model could have resulted in gaseous regions at a range of radii, including those observed here. 

\subsection{X-ray ionized nova ejecta} 

Another potential source of the nebular gas is the ejecta from a nova in the cluster \citep{Ripamonti12}. In this model, the accreting object photoionizes the ejecta from an unrelated nova in the same cluster. Because the ionizing source and nebula gas are unrelated in this model, the accreting compact object may have any mass, so long as it produces sufficient X-ray luminosity. \citet{Ripamonti12} demonstrated that this model is capable of reproducing both the [O~{\sc iii}] doublet and the lack of other emission lines in the observed optical spectrum of RZ~2109. However, their model requires that the nova ejecta shell is $\lesssim$0.1~pc from the X-ray source. This is significantly smaller than the observed nebula, which is found to be on parsec scales. 

\subsection{IMBH accretion disk origin} 

Previous studies have considered the possibility that the large velocity width of the [O~{\sc iii}]~$\lambda$5007 emission is due to gravitational motions around a BH. \citet{Zepf08} found that such a model fails because, at the radius where such velocities are available, there is not enough volume to produce the observed [O~{\sc iii}] luminosity (given its critical density) unless the mass of the BH is a significant fraction of the total GC mass. \citet{Porter10} confirmed this calculation, and showed specifically that such a scenario was only possible with a BH mass $>$30,000~M$_{\odot}$. If the width of the [O~{\sc iii}] emission is due to Keplerian motions around a 30,000~M$_{\odot}$ BH, then the emitting region would be around 10$^{-6}$~pc. Such a region is many orders of magnitude beneath the spatial resolution of our observations and would appear unresolved. Given that the [O~{\sc iii}] emission line is resolved, these observations rule out such an origin. 

\subsection{Tidal disruption of a white dwarf/ horizontal branch star by an IMBH}

Another model, invoked to explain ultraluminous X-ray emission and optical emission lines, is the tidal disruption of a white dwarf by a central IMBH \citep[e.g.,][]{Irwin10}. In this model, debris from the tidally disrupted star is accreted on to the central IMBH. The resulting outburst can then photoionize the remaining debris, hence producing both the observed X-ray and nebular emission. \citet{Clausen11} noted that, while such a model can produce both the X-ray emission and the luminosity and width of the [O~{\sc iii}]~$\lambda$5007 emission line observed from RZ~2109, the emission should peak after a few years. Such a timescale is not long enough to explain the observed X-ray emission from the cluster, which suggests that the source has been in outburst for over 14~yrs. It is also inconsistent with the nebular emission observed here, whose scale of $\gtrsim$~2~pc and velocity of 1500~kms$^{-1}$ suggest a timescale of $\gtrsim$~1300~yrs.  

\citet{Clausen12} extended this work to consider the tidal disruption of a horizontal branch star by an IMBH. They demonstrated that this could produce the X-ray, [O~{\sc iii}] and [N~{\sc ii}] emission observed from another GC BH candidate, the X-ray source CXOJ033831.8352604 in an NGC~1399 GC. The disruption of a horizontal branch star predicts a longer outburst timescale with $\tau\sim$200~yrs \citep{Clausen12} and is therefore capable of producing the observed X-ray emission. At the observed velocity of the emission line in RZ~2109, $\sigma\sim$1500~kms$^{-1}$ \citep{Zepf08}, the maximum associated nebula scale is simply $\tau\times\sigma\lesssim0.3$~pc. Such a nebula would be unresolved by these observations and is smaller than the best fitting half light radii to the observed [O~{\sc iii}] emission. Therefore, the tidal disruption of a star by an IMBH is unlikely to be the source of the emission observed in RZ~2109.

\section{Conclusions}

The STIS spectroscopy presented in this paper spatially resolves both the stellar and nebular emission from the GC RZ~2109. By considering the [O~{\sc iii}]~$\lambda\lambda$4959,~5007${\rm \AA}$ emission lines, we have shown that the nebula in this cluster is extended, with best fitting half light radii in the range 3-7~pc. 

The observed size of this nebula is larger than that predicted by several models that have been proposed to explain the nebular and ULX emission detected from this cluster. It is many orders of magnitude larger than that expected from emission from an accretion disk around an IMBH (which should be on scales of 10$^{-6}$~pc). Similarly, the tidal disruption of a WD or HB star by an IMBH predicts emission on timescales that produce nebulae that are an order of magnitude smaller than observed (0.05~pc and 0.3~pc, respectively). The nebula is also an order of magnitude larger than that predicted from the ionization of ejecta from a nova in the cluster, which should produce nebulae on scales of 0.1~pc. 

The large nebula observed could be produced via a wind driven from a BH accreting at a rate close to the Eddington limit. The timescales for such accretion are unknown. However, it is possible that such a system has been accreting for long enough (either persistently or with transient behavior) to drive the nebula gas to the few pc scales observed. 

\section*{Acknowledgements}
We thank the anonymous referee for considering this paper and for their helpful comments. We are grateful for the assistance of the {\it HST} STIS helpdesk team, who provided very useful answers and suggestions in relation to the analysis of these data. Support for this work was provided by NASA through grant numbers HST-GO-11703 (MBP and SEZ) and HST-GO-11029 (SEZ and CZW) from the Space Telescope Science Institute, which is operated by AURA, Inc., under NASA contract NAS 5-26555. Support for AK's work was provided by NASA through Chandra grant numbers GO1-12112X and GO0-11111A issued by the Chandra X-ray Observatory Center, which is operated by the SAO for and on behalf of the NASA under contract NAS8-03060. KLR is supported by an NSF Faculty Early Career Development (CAREER) award (AST-0847109). The work of DS was carried out at Jet Propulsion Laboratory, California Institute of Technology, under a contract with NASA.

\bibliographystyle{apj_w_etal}
\bibliography{bibliography_etal}

\begin{thebibliography}{40}
\expandafter\ifx\csname natexlab\endcsname\relax\def\natexlab#1{#1}\fi

\bibitem[{{Angelini} {et~al.}(2001){Angelini}, {Loewenstein}, \&
  {Mushotzky}}]{Angelini01}
{Angelini}, L., {Loewenstein}, M., \& {Mushotzky}, R.~F. 2001, ApJ, 557, L35

\bibitem[{{Barnard} {et~al.}(2012){Barnard}, {Garcia}, \& {Murray}}]{Barnard12}
{Barnard}, R., {Garcia}, M., \& {Murray}, S.~S. 2012, ArXiv e-prints

\bibitem[{{Barnard} {et~al.}(2008){Barnard}, {Stiele}, {Hatzidimitriou},
  {Kong}, {Williams}, {Pietsch}, {Kolb}, {Haberl}, \& {Sala}}]{Barnard08}
{Barnard}, R. {et~al.} 2008, \apj, 689, 1215

\bibitem[{{Bostroem} \& {Proffitt}(2011)}]{Bostroem11}
{Bostroem}, K. \& {Proffitt}, C. 2011, {STIS Data Handbook, Version 6.0}
  (STScI: Baltimore)

\bibitem[{{Brassington} {et~al.}(2010){Brassington}, {Fabbiano}, {Blake},
  {Zezas}, {Angelini}, {Davies}, {Gallagher}, {Kalogera}, {Kim}, {King},
  {Kundu}, {Trinchieri}, \& {Zepf}}]{Brassington10}
{Brassington}, N.~J. {et~al.} 2010, \apj, 725, 1805

\bibitem[{{Brassington} {et~al.}(2012){Brassington}, {Fabbiano}, {Zezas},
  {Kundu}, {Kim}, {Fragos}, {King}, {Pellegrini}, {Trinchieri}, {Zepf}, \&
  {Wright}}]{Brassington12}
---. 2012, \apj, 755, 162

\bibitem[{{Carlson} \& {Holtzman}(2001)}]{Carlson01}
{Carlson}, M.~N. \& {Holtzman}, J.~A. 2001, PASP, 113, 1522

\bibitem[{{Clark}(1975)}]{Clark75}
{Clark}, G.~W. 1975, ApJ, 199, L143

\bibitem[{{Clausen} \& {Eracleous}(2011)}]{Clausen11}
{Clausen}, D. \& {Eracleous}, M. 2011, \apj, 726, 34

\bibitem[{{Clausen} {et~al.}(2012){Clausen}, {Sigurdsson}, {Eracleous}, \&
  {Irwin}}]{Clausen12}
{Clausen}, D., {Sigurdsson}, S., {Eracleous}, M., \& {Irwin}, J.~A. 2012, ArXiv
  e-prints

\bibitem[{{Fabbiano}(2006)}]{Fabbiano06}
{Fabbiano}, G. 2006, ARA\&A, 44, 323

\bibitem[{{Goudfrooij} {et~al.}(2006){Goudfrooij}, {Bohlin},
  {Ma{\'{\i}}z-Apell{\'a}niz}, \& {Kimble}}]{Goudfrooij06}
{Goudfrooij}, P., {Bohlin}, R.~C., {Ma{\'{\i}}z-Apell{\'a}niz}, J., \&
  {Kimble}, R.~A. 2006, \pasp, 118, 1455

\bibitem[{{Hurley}(2007)}]{Hurley07}
{Hurley}, J.~R. 2007, \mnras, 379, 93

\bibitem[{{Irwin} {et~al.}(2010){Irwin}, {Brink}, {Bregman}, \&
  {Roberts}}]{Irwin10}
{Irwin}, J.~A., {Brink}, T.~G., {Bregman}, J.~N., \& {Roberts}, T.~P. 2010,
  \apjl, 712, L1

\bibitem[{{Jord{\'a}n} {et~al.}(2004)}]{Jordan04}
{Jord{\'a}n}, A. {et~al.} 2004, ApJ, 613, 279

\bibitem[{{Jord{\'a}n} {et~al.}(2007)}]{Jordan07}
---. 2007, ApJ, 671, L117

\bibitem[{{Kalogera} {et~al.}(2004){Kalogera}, {King}, \& {Rasio}}]{Kalogera04}
{Kalogera}, V., {King}, A.~R., \& {Rasio}, F.~A. 2004, \apjl, 601, L171

\bibitem[{{Kim} {et~al.}(2006){Kim}, {Kim}, {Fabbiano}, {Lee}, {Park},
  {Geisler}, \& {Dirsch}}]{Kim06}
{Kim}, E., {Kim}, D., {Fabbiano}, G., {Lee}, M.~G., {Park}, H.~S., {Geisler},
  D., \& {Dirsch}, B. 2006, ApJ, 647, 276

\bibitem[{{King} {et~al.}(1996){King}, {Kolb}, \& {Burderi}}]{King96}
{King}, A.~R., {Kolb}, U., \& {Burderi}, L. 1996, ApJ, 464, L127+

\bibitem[{{Kundu} {et~al.}(2007){Kundu}, {Maccarone}, \& {Zepf}}]{Kundu07}
{Kundu}, A., {Maccarone}, T.~J., \& {Zepf}, S.~E. 2007, ApJ, 662, 525

\bibitem[{{Maccarone} {et~al.}(2007){Maccarone}, {Kundu}, {Zepf}, \&
  {Rhode}}]{Maccarone07}
{Maccarone}, T.~J., {Kundu}, A., {Zepf}, S.~E., \& {Rhode}, K.~L. 2007, \nat,
  445, 183

\bibitem[{{Maccarone} {et~al.}(2010){Maccarone}, {Kundu}, {Zepf}, \&
  {Rhode}}]{Maccarone10}
---. 2010, \mnras, 409, L84

\bibitem[{{Maccarone} {et~al.}(2011){Maccarone}, {Kundu}, {Zepf}, \&
  {Rhode}}]{Maccarone11a}
---. 2011, \mnras, 410, 1655

\bibitem[{{Madrid} {et~al.}(2009){Madrid}, {Harris}, {Blakeslee}, \&
  {G{\'o}mez}}]{Madrid09}
{Madrid}, J.~P., {Harris}, W.~E., {Blakeslee}, J.~P., \& {G{\'o}mez}, M. 2009,
  \apj, 705, 237

\bibitem[{{Maraston}(2005)}]{Maraston05}
{Maraston}, C. 2005, MNRAS, 362, 799

\bibitem[{{Peacock} {et~al.}(2009){Peacock}, {Maccarone}, {Waters}, {Kundu},
  {Zepf}, {Knigge}, \& {Zurek}}]{Peacock09}
{Peacock}, M.~B., {Maccarone}, T.~J., {Waters}, C.~Z., {Kundu}, A., {Zepf},
  S.~E., {Knigge}, C., \& {Zurek}, D.~R. 2009, MNRAS, 392, L55

\bibitem[{{Peacock} {et~al.}(2012){Peacock}, {Zepf}, \&
  {Maccarone}}]{Peacock12}
{Peacock}, M.~B., {Zepf}, S.~E., \& {Maccarone}, T.~J. 2012, \apj, 752, 90

\bibitem[{{Porter}(2010)}]{Porter10}
{Porter}, R.~L. 2010, \mnras, 407, L59

\bibitem[{{Proga}(2007)}]{Proga07}
{Proga}, D. 2007, \apj, 661, 693

\bibitem[{{Ripamonti} \& {Mapelli}(2012)}]{Ripamonti12}
{Ripamonti}, E. \& {Mapelli}, M. 2012, \mnras, 2958

\bibitem[{{Sarazin} {et~al.}(2000){Sarazin}, {Irwin}, \& {Bregman}}]{Sarazin00}
{Sarazin}, C.~L., {Irwin}, J.~A., \& {Bregman}, J.~N. 2000, \apjl, 544, L101

\bibitem[{{Sarazin} {et~al.}(2003){Sarazin}, {Kundu}, {Irwin}, {Sivakoff},
  {Blanton}, \& {Randall}}]{Sarazin03}
{Sarazin}, C.~L., {Kundu}, A., {Irwin}, J.~A., {Sivakoff}, G.~R., {Blanton},
  E.~L., \& {Randall}, S.~W. 2003, ApJ, 595, 743

\bibitem[{{Shih} {et~al.}(2010){Shih}, {Kundu}, {Maccarone}, {Zepf}, \&
  {Joseph}}]{Shih10}
{Shih}, I.~C., {Kundu}, A., {Maccarone}, T.~J., {Zepf}, S.~E., \& {Joseph},
  T.~D. 2010, \apj, 721, 323

\bibitem[{{Steele} {et~al.}(2011){Steele}, {Zepf}, {Kundu}, {Maccarone},
  {Rhode}, \& {Salzer}}]{Steele11}
{Steele}, M.~M., {Zepf}, S.~E., {Kundu}, A., {Maccarone}, T.~J., {Rhode},
  K.~L., \& {Salzer}, J.~J. 2011, \apj, 739, 95

\bibitem[{{Trenti} {et~al.}(2007){Trenti}, {Ardi}, {Mineshige}, \&
  {Hut}}]{Trenti07}
{Trenti}, M., {Ardi}, E., {Mineshige}, S., \& {Hut}, P. 2007, \mnras, 374, 857

\bibitem[{{Verbunt} \& {Hut}(1987)}]{Verbunt87}
{Verbunt}, F. \& {Hut}, P. 1987, in IAU Symposium, Vol. 125, The Origin and
  Evolution of Neutron Stars, ed. {D.~J.~Helfand \& J.-H.~Huang}, 187--+

\bibitem[{{Waters}(2007)}]{Waters07}
{Waters}, C.~Z. 2007, PhD thesis, AA(Michigan State University)

\bibitem[{{Williams} {et~al.}(2007){Williams}, {Ciardullo}, {Durrell},
  {Feldmeier}, {Sigurdsson}, {Vinciguerra}, {Jacoby}, {von Hippel}, {Ferguson},
  {Tanvir}, {Arnaboldi}, {Gerhard}, {Aguerri}, \& {Freeman}}]{Williams07}
{Williams}, B.~F. {et~al.} 2007, \apj, 654, 835

\bibitem[{{Zepf} {et~al.}(2007){Zepf}, {Maccarone}, {Bergond}, {Kundu},
  {Rhode}, \& {Salzer}}]{Zepf07}
{Zepf}, S.~E., {Maccarone}, T.~J., {Bergond}, G., {Kundu}, A., {Rhode}, K.~L.,
  \& {Salzer}, J.~J. 2007, \apjl, 669, L69

\bibitem[{{Zepf} {et~al.}(2008){Zepf}, {Stern}, {Maccarone}, {Kundu},
  {Kamionkowski}, {Rhode}, {Salzer}, {Ciardullo}, \& {Gronwall}}]{Zepf08}
{Zepf}, S.~E. {et~al.} 2008, \apjl, 683, L139

\end{thebibliography}

\label{lastpage}

\end{document}